\documentclass[sigconf]{acmart}

\AtBeginDocument{%
  \providecommand\BibTeX{{%
    Bib\TeX}}}

\copyrightyear{2022} 
\acmYear{2022} 
\setcopyright{rightsretained} 
\acmConference[RecSys '22]{Sixteenth ACM Conference on Recommender
Systems}{September 18--23, 2022}{Seattle, WA, USA}
\acmBooktitle{Sixteenth ACM Conference on Recommender Systems
(RecSys '22), September 18--23, 2022, Seattle, WA, USA}
\acmDOI{10.1145/3523227.3547394}
\acmISBN{978-1-4503-9278-5/22/09}

\begin{document}

\newcommand{\jiajing}[1]{{{\textcolor{red}{[Jiajing: #1]}}}}
\newcommand{\longpsnew}{PinnerFormer}
\newcommand{\longpsold}{PinnerSage}
\newcommand{\psnew}{PinnerFormer}

\title{Rethinking Personalized Ranking at Pinterest: An End-to-End Approach}

\author{Jiajing Xu}
\email{jiajing@pinterest.com}
\affiliation{%
  \institution{Pinterest Inc.}
  \city{San Francisco}
  \country{USA}
}

\author{Andrew Zhai}
\email{andrew@pinterest.com}
\affiliation{%
  \institution{Pinterest Inc.}
  \city{San Francisco}
  \country{USA}
\email{andrew@pinterest.com}
}

\author{Charles Rosenberg}
\email{crosenberg@pinterest.com}
\affiliation{%
  \institution{Pinterest Inc.}
  \city{San Francisco}
  \country{USA}
}

\renewcommand{\shortauthors}{Xu et al.}

\begin{abstract}
In this work, we present our journey to revolutionize the personalized recommendation engine through end-to-end learning from raw user actions.  We encode user's long-term interest in \longpsnew{}, a user embedding optimized for long-term future actions via a new dense all-action loss, and capture user's short-term intention by directly learning from the real-time action sequences.  We conducted both offline and online experiments to validate the performance of the new model architecture, and also address the challenge of serving such a complex model using mixed CPU/GPU setup in production.  The proposed system has been deployed in production at Pinterest and has delivered significant online gains across organic and Ads applications.

\end{abstract}



\maketitle

\section{Introduction}

Pinterest's mission is to bring over 400M monthly active users the inspiration to create a life they love.  On each major surface at Pinterest (Homefeed, Related Pins and Search), we generate personalized recommendations based on user's interaction, such as saving Pins to board (repin), clicking through to the underlying link, zooming in on one Pin (close-up), hiding irrelevant content, and more.  

A better understanding of the users is crucial to provide such personalization recommendations.  It is very common across the industry to learn user embeddings to power personalized recommendations  \cite{covington2016deep,yang2020mixed,grbovic2018realtime,zhang2020personalized,li2019multiinterest, pi2019practice}.  One alternative that has shown promise is to directly build ranking models to generate personalized recommendations using sequential information from a user's recent engagement \cite{chen2019behavior, zhou2019deep, pi2019practice, pi2020searchbased}.  However, sequential models traditionally focus on a real-time setting, aiming to predict the next immediate action from all prior actions.  There is limited work on leveraging the complete user action sequences \cite{fazelnia2022variational}.  In this work, we decompose user intentions into long-term interest and short-term intention, and optimize for long-term future actions and immediate next action respectively.

\section{Methods}
\begin{figure*}
    \centering
    \includegraphics[width=0.8\textwidth]{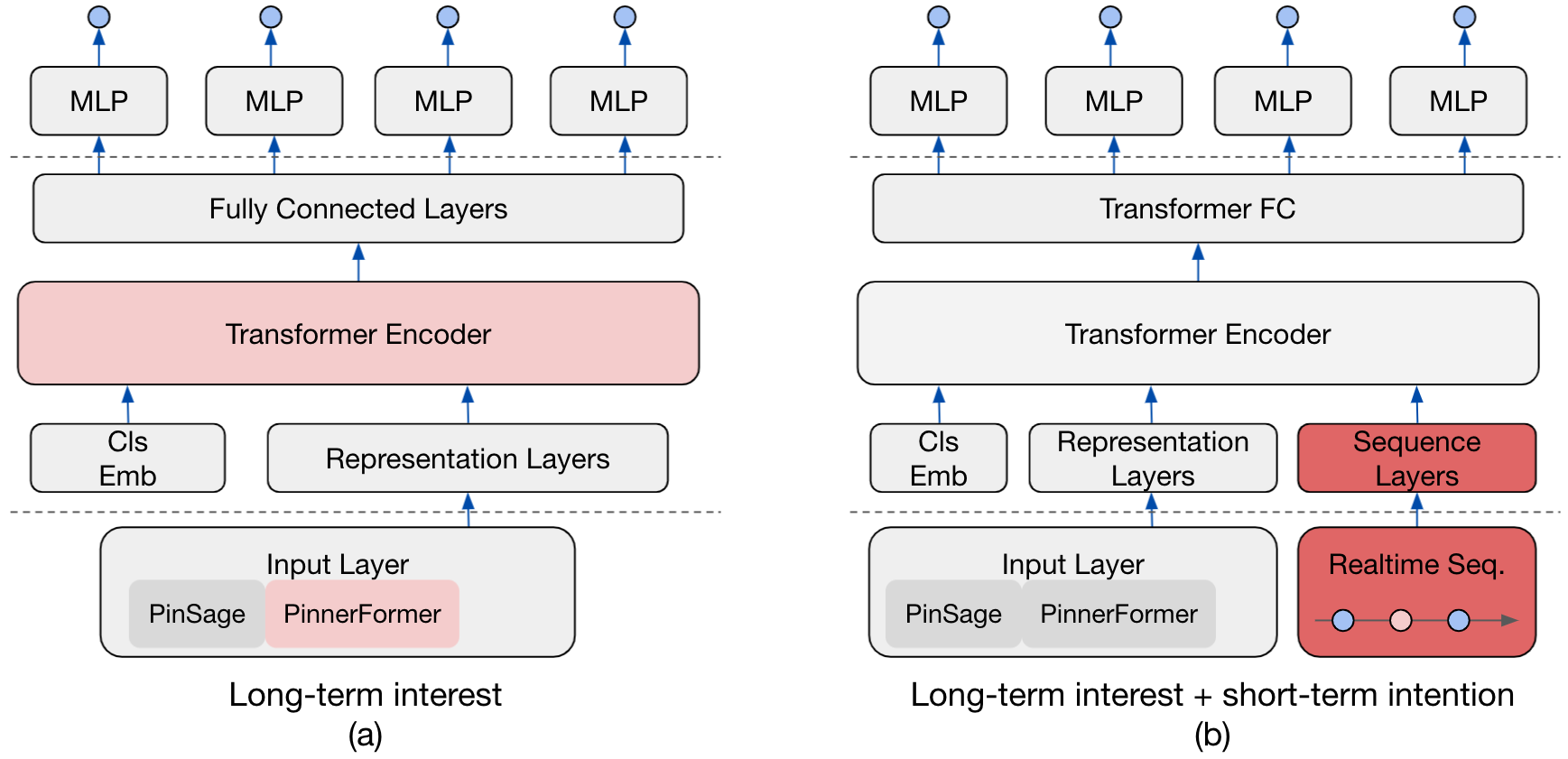}
    \caption{System architecture (a) long-term interest only and (b) long-term interest and short-term intention.}
    \label{fig:system_arch}
\end{figure*}

Figure \ref{fig:system_arch} shows the overall model architecture.  We will talk about the two new components in more details below.

\subsection{Encode Long-term User Interest -- \longpsnew{}}
We introduce \longpsnew{}, an end-to-end learned user embeddings via sequence modeling to encode long-term user interest \cite{pancha2022pinnerformer}.  A key contribution of this work is to learn a model that is able to predict a user's \textit{positive} future engagement over a 14-day time window after the generation of their embedding, rather than a traditional sequence modeling task, where the embedding would only predict the next action taken.  We choose this range of 14 days for tractability, and assume that actions a user takes over the course of two weeks sufficiently are representative of a user's longer-term interest.  Figure \ref{fig:e2e_model} illustrates the \longpsnew{} architecture, and below we briefly go over each component.  

\subsubsection{Feature Representation}
Each Pin at Pinterest is represented with a PinSage \cite{ying2018graph} embedding, which is an aggregation of visual \cite{10.1145/2783258.2788621,10.1145/3366423.3380031}, text annotations \cite{NIPS2013_9aa42b31, NIPS2016_c24cd76e}, and engagement information \cite{10.1145/3178876.3186183} for a Pin.

For each user, we collect a sequence of actions a user has taken on Pinterest over the past year ordered ascending by timestamp, where we limit the actions to users' engagements with Pins, including Pin saves, clicks, reactions, and comments.  An action can then be represented by a PinSage embedding, and other metadata features: action type, timestamp, action duration, and surface.  All features are concatenated into a single vector.  To keep the problem tractable, we compute a user's embedding using their $M$ most recent actions.

\subsubsection{Model Architecture}
In \longpsnew{}, we model the sequence of user actions using a transformer model architecture \cite{vaswani2017attention}.
We choose to use PreNorm residual connections, applying Layer Normalization before each block, as this approach has been shown to improve stability of training \cite{nguyen2019transformers, wang2019learning}.

On the User side, we put together the input matrix using the $M$ actions leading up to the next action in the user's sequence.  Then, we project the input to the transformer's hidden dimension, add a fully learnable positional encoding, and apply a standard transformer consisting of alternating feedforward network (FFN) and multi-head self attention (MHSA) blocks.
The output of the transformer at every position is passed through a small MLP and $L_2$ normalized, resulting in a set of the final embedding.

On the Pin side, we learn an MLP that takes only PinSage as an input, and $L_2$ normalize the output embeddings.


\subsubsection{Training Objective}
\begin{figure*}
    \centering
    \includegraphics[width=0.8\textwidth]{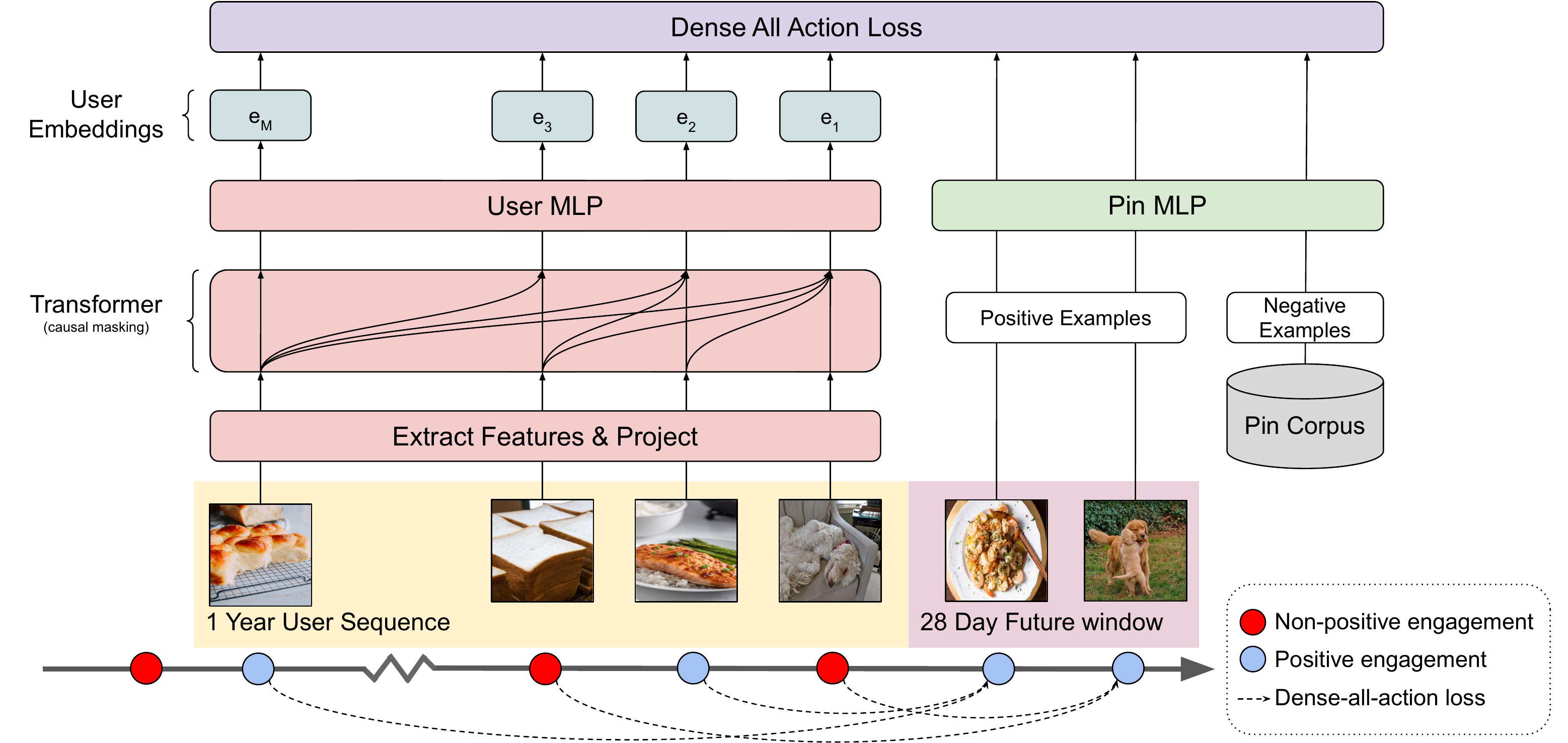}
    \caption{Overview of \psnew{} architecture. Features are passed through a transformer with causal masking, and embeddings are returned at every time step.
    Blue circles represent embeddings associated with positive actions, while red circles represent embeddings associated with non-positive actions (but not necessarily negative) 
    }
    \label{fig:e2e_model}
\end{figure*}

We evaluated many methods of structuring our learning objective and introduce the new “Dense All Action” loss, shown in Figure \ref{fig:e2e_model}. Specifically given a user embedding we predict all positive actions in the next 28 days, averaging each action contribution. We call this a “Dense” loss because we apply this to every output of the TransformerEncoder where given $M$ inputs to the transformer we can product $M$ outputs. With a causal masking of the attention weights we have $M$ user embeddings to apply the all positive action loss.

\subsection{Capture Short-term User Intention -- Real-time User Sequences}
We collect a user's past $P$ actions from real-time logging to capture user's short-term intention.  Each action is represented by the following features: action timestamp, action duration, action types (e.g. repin, hide), and PinSage embeddings. These features are then passed through a MHSA block, and fed into the TransformerEncoder layer along with the \longpsnew{} embedding.

In our previous attempt to include real-time action features, the model become too responsive to user recent action (e.g. after a user engages with a Pin from food category, the next refresh of the Homefeed page is filled with more food Pins), which is not a desired user experience.  In order to mitigate such behavior, we added time window masks to the real-time action sequences during training and successfully reduced the model sensitivity.


\subsection{Model Serving}
The proposed model architecture is undoubtedly more powerful than previous models served in production.  This comes at a high price - increased infrastructure cost and serving latency, which are hard blockers in real-world deployment.  We had no luck in serving the model on most powerful CPU on AWS (300\% increase in latency!), thus we turned our focus to moving some parts of the model operations to GPU serving. Specifically while ranking models have hundreds of small ops coming from hundreds of input features, we can move parts of the computation ( e.g. TransformerEncoder) to GPU to speedup computation. 

Table \ref{tab:serving} illustrates how each optimization improved latency.  Initially we saw that moving the CPU model entirely to GPU increases latency, but by selecting which operations and layers are on what device we are able to substantially reduce the latency to the point where GPU with a Transformer versus CPU without the Transformer have nearly similar latency and cost.  This allowed us to continue online experiments and ship the model in production eventually.

\begin{table}
    \centering
    \caption{Latency increase to serve the proposed model on (a) CPU only (b) GPU only (c) Mixed CPU-GPU where embedding lookup is done on CPU (d) Mixed CPU-GPU where embedding lookup and projection is done on CPU (e) Mixed CPU-GPU where embedding lookup, projection, and feature pre-processing is done on CPU.}
    \label{tab:serving}
    
    \begin{tabular}{lr}
        \toprule
        Serving Method & Latency Increase  \\
        \midrule
        (a) CPU Model & +300\%  \\
        (b) GPU Model & +400\% \\
        (c) Consolidated Ops (Embeddings)  & +250\% \\
        (d) Consolidated Ops (Projections) & +150\% \\
        (e) Mixed Device  & +10\% \\
        \bottomrule
    \end{tabular}
\end{table}

\begin{table}
    \centering
    \caption{Online A/B experiment results. We see improvements in both organic and Ads metrics.}
    \label{tab:pbty-exp}
    
    \begin{tabular}{c@{\qquad}c}
    \begin{tabular}{lcc}
        \toprule
        Model  & \begin{tabular}{@{}c@{}}Lift\\(Organic Repin)\end{tabular} & \begin{tabular}{@{}c@{}}Lift\\(Ads CTR)\end{tabular} \\
        
        \cmidrule(r){1-1}\cmidrule(l){2-3}
        \longpsnew{}  & +7.5\% & +9.1\% \\
        \begin{tabular}{@{}c@{}}\longpsnew+Real-time\end{tabular}  & +12.5\%  & +14.0\% \\
        \bottomrule
    \end{tabular}
    \end{tabular}
\end{table}

\section{Experiments and Results}
We conducted extensive experiments to evaluate the performance of the proposed model architecture in both offline and online settings.  In our offline evaluations, we observed superior performance compared to the previous baseline \cite{pancha2022pinnerformer} and tuned the parameters to strike a good balance between infrastructure cost and performance gains, i.e. $M$=255, $P$=100.  
Additionally, we run several A/B experiments to better understand how the systems perform online.

\subsection{Homefeed Ranking}
We first run experiment on Pinterest's Homefeed ranking model, which ranks the order of Pins to be shown on a user's Homefeed.
In the control group, the previous production model used a weighted average of a user's top $k$ \longpsold{} embeddings as a feature \cite{pal2020pinnersage}.
In one treatment group, we replaced this aggregation of \longpsold{} with the single \psnew{} embedding.  In another treatment group, we further added the real-time action sequence.  The ranking models across all the experiment groups are trained on the same date range of data for fair comparisons.

As shown in Table \ref{tab:pbty-exp}, \psnew{} significantly improved organic engagement (repins) on Homefeed, while the real-time action sequence complements \psnew{} and lifted the metrics further.

\subsection{Related Pins Ads Ranking}
To test the generalization of this model architecture, we run an A/B experiment on the Related Pins Ads ranking model, which ranks the order of advertisement Pins to be shown to the user's Related Pins feed.  Due to some infrastructure constraints at the time of publication, we replaced PinSage embeddings with Pin text annotations to represent each user action in the real-time action sequence.  The previous production model used a slightly different method to aggregate user's top embeddings.  The treatment groups followed the similar setup as in the above experiment.  

In Table \ref{tab:pbty-exp}, we observe consistent results where both treatment groups show significant gains in engagement with Ads measured by click-through rate (CTR).

\section{Conclusion}
Capturing both long-term user interest and short-term user intention has greatly improved the personalization experience on Pinterest.  In the future, we plan to further expand the user actions to be included in the sequence of user actions, and explore this model architecture in candidate generation stage.

\begin{acks}
We thank our colleagues
Jay Adams, 
Dhruvil Badani, 
Kofi Boakye, 
Haoyu Chen, 
Ludek Cigler, 
Pong Eksombatchai, 
Nazanin Farahpour, 
Neng Gu, 
Yi-ping Hsu, 
Jure Leskovec, 
Haomiao Li, 
Yang Liu, 
Cosmin Negruseri, 
Nikil Pancha, 
Yan Sun, 
Po-wei Wang, 
Xue Xia, 
Zhiyuan Zhang, 
and Yitong Zhou for their help and support with the project.
\end{acks}

\section*{Speaker Bio}
\textbf{Jiajing Xu} is a Senior Machine Learning Engineering Manager at Pinterest, where he leads the Applied Research teams across Representation Learning, Personalization and Inclusive AI.  Prior to his current role, he was the first computer vision scientist at Pinterest and co-founded the visual discovery team, created and grew Related Pins Ads product, and managed the Ads Ranking team.  He holds a Ph.D. and a Master's degree from Stanford University, and a Bachelor's degree from California Institute of Technology.

\bibliographystyle{ACM-Reference-Format}
\bibliography{recsys2022}

\end{document}